\newcommand{\beq}{\begin{equation}}
\newcommand{\eeq}{\end{equation}}
\newcommand{\bea}{\begin{eqnarray}}
\newcommand{\eea}{\end{eqnarray}}

\def\pmu{\partial_{\mu}}
\def\pamu{\partial^{\mu}}

\documentstyle[12pt]{article}

\title{Collective Coordinate Quantization:\\ Relativistic
and Gauge Symmetric Aspects}
\author{R. Sollacher\thanks{E-mail address: sollach@tpri6c.gsi.de}
\\ GSI, P.O.Box 110552, D-64220 Darmstadt, Germany}

\begin{document}

\maketitle

\abstract{The introduction and quantization of a center-of-mass
coordinate is demonstrated for the one-soliton sector of nonlinear
field theories in (1+1) dimensions. The present approach strongly
emphazises the gauge and BRST-symmetry aspects of collective
coordinate quantization. A gauge is presented which is independent of
any approximation scheme and which allows to interpret the new degree
of freedom as the {\em quantized} center of mass coordinate of a
soliton. Lorentz invariance is used from the beginning to introduce
fluctuations of the collective coordinate in the {\em rest frame} of
the {\em moving} soliton. It turns out that due to the extended nature
of the soliton retardation effects lead to differences in the quantum
mechanics of the soliton as compared to a point-like particle.
Finally, the results of the semiclassical expansion are used to
analyse effective soliton-meson vertices and the coupling to an
external source.  Such a coupling in general causes acceleration as
well as internal excitation of the soliton.}

\vspace{-195mm}
\vfill\hfill
\vbox{
\hfill GSI-93-56\null\par
\hfill hep-th/9307098}\null
\vspace{195mm}

\clearpage

\section{Introduction}

It is a well known fact that the choice of a ground state violating
certain symmetries of the underlying theory leads to the appearance of
zero modes after quantization.  Prominent examples are mean field
systems like in nuclear physics \cite{Ring} or soliton systems
\cite{Raj}. In these cases the ground state,
represented by the mean field respectively the classical soliton,
violates certain symmetries, among them translational symmetry.  The
associated zero energy modes are by no means unphysical: their
excitation, which costs no energy, corresponds to infinitesimal
translations and eventually leads to a spread of the wave function
over the whole space such that translational symmetry is restored.  In
the ideal case one recovers an eigenstate of the momentum operator in
analogy to plane waves.

Why does one bother about these zero modes? One reason is that they
cause problems in numerical calculations due to their singular
behaviour in the infrared limit. Another reason is a physical one:
Whereas in the mathematical description these zero modes are just
modes like all others, except for their zero energy eigenvalue, they
are well separated from the other, non-zero modes in physical
pictures; the zero modes correspond to collective motion, whereas the
other modes are identified with intrinsic excitations. As such the
zero modes are of quite different nature.  Therefore, it appears
desirable to represent the zero modes by suitable collective
coordinates in contrast to the intrinsic modes. Once this is achieved
one also has a better handle on the intrinsic modes eventually free of
infrared problems.

In the following I am concentrating on nonlinear field theories in
(1+1) dimensions possessing soliton solutions like the Sine-Gordon or
$\phi^4$-model. There, the only zero mode is due to translational
motion. Soon after the first attempts on quantizing the soliton sector
dating back to the mid-seventies
\cite{Sol,GJ,CL} (for a review see also \cite{Raj}) a lot of
attention has been attracted by the zero-mode problem
\cite{CL,GS,CC,ZM,FK,Ab}. One particular route approaching this
problem \cite{CL,GS,CC} is based mainly on the collective coordinate
method in many-body theory \cite{BoPi}. Another topic addressed in
connection with translational motion is Lorentz covariance. Using
either an explicitly Lorentz covariant approach or summing up an
infinite series of Feynman diagrams Lorentz covariance has been
demonstrated \cite{CL,GS,FK,Rel}.

Also in the midseventies it has been realized that there exists a
connection between the introduction of collective coordinates and
gauge theories \cite{HoKi}. Only within the last few years this
connection has been investigated closer and finally led to a
BRST-symmetric treatment of collective coordinate quantization
\cite{BRSTqu}. The method presented in the following strongly
emphazises the gauge- and BRST-symmetry aspects of collective
coordinate quantization. In contrast to most other approaches I am
using a Lagrangian formulation based on a functional integral
representation of the time evolution operator. The principles of the
present approach are described extensively in a work by J.  Alfaro and
P.H. Damgaard \cite{AlDam}. The following treatment of the
center-of-mass motion of an extended particle can be regarded as an
application of their ideas.

Apart from these more technical aspects of collective coordinate
quantization special attention will be paid to fluctuations of the
collective coordinate. These fluctuations are responsible for a
diffusion of the soliton's position, $i.e.$ the probability to find
the soliton at a certain position spreads over the whole space. At
first glance the relativistic covariant expressions for the soliton's
momentum and energy suggest the picture of a relativistic point-like
particle. However, such a picture certainly contradicts causality; as
an extended object the soliton must be subject to retardation effects
as soon as it becomes accelerated. Fluctuations of the center-of-mass
coordinate just provide such an acceleration. Related critisism in
this direction has been presented in \cite{Ral}. It will turn out that
precisely due to retardation effects fluctuations of the collective
coordinate of a soliton behave different as those of a point-like
particle.

Before proceeding with the details let me present an outline of what
follows. Section 2 gives a short review about soliton quantization as
far as it concerns the present subject. In section 3 I explain the
introduction of a collective coordinate and its relation to gauge
theories. The r\^ole of BRST-symmetry in connection with approximation
schemes is demonstrated with the example of a naive semiclassical
expansion.  Section 4 describes a slightly different treatment of the
collective fluctuations requiring a formal solution of the gauge
constraint.  Within a tree approximation the difference to the quantum
mechanics of a relativistic point-particle is shown. In section 5 I
address the question of soliton-meson vertices and the effect of an
external source. Finally, section 6 contains a summary, a discussion
of the results and an outlook concerning an extension to (3+1)
dimensions.

\section{Soliton quantization: A short review}

The starting point will be the matrix element of the time evolution
operator between states $\langle f|$ and $|i\rangle$ using the
functional intergral representation:
\bea
{\cal Z} [j] &=& \langle f| {\cal T} \exp \left( -i\int_{-T'/2}^{T'/2}
dt \hat{H} (t) \right) |i \rangle \cr &=&
\int\! {\cal D}[\phi ] \;
\exp\left( i\int\! dxdt\; {\cal L}_g (x,t)\right)
\eea
Here, $\hat{H} (t)$ is the hamilton operator related to the Lagrangian
\beq
{\cal L}_g = \frac{1}{2} \pmu \phi \pamu \phi - U(\phi ,g) + j
\phi
\eeq
with $\phi (x,t)$ a dimensionless scalar field. Functional integration
is to be taken over all configurations $\phi (x,t)$ with boundary
conditions in time given by $\langle f|$ and $|i\rangle$. The
dimensionless coupling constant $g$ measures the strength of a
nonlinear selfinteraction $U(\phi ,g)$. Examples are the
$\phi^4$-theory or the Sine-Gordon-theory,
\beq
U_{\phi^4}(\phi ,g) = m^2 \phi^2 (g^2\phi^2 -1)/2~~,~~ U_{SG}(\phi ,g)
= m^2(1- \cos g\phi)/g^2~~,
\eeq
both exhibiting soliton solutions. It turns out useful to rescale
$\phi \to \phi /g$ thus explicitly demonstrating that $g^2$ plays the
same r\^ole as $\hbar$; therefore, the latter can be absorbed into
$g^2$. This leads to
\bea
{\cal Z} [j] &=& \int\! {\cal D}[\phi ] \; \exp \left(
\frac{i}{g^2}\int\!  dxdt\; {\cal L}(x,t) \right) \cr {\cal L} &=&
\frac{1}{2} \pmu \phi \pamu \phi - V(\phi ) + g j \phi
\label{eq:genf}
\eea
with $V(\phi ) = U(\phi , g=1)$. The external source $j(x,t)$ can be
viewed as inducing weak perturbations thus probing the response of the
system.

A common feature of solitonic sytems is a degeneracy of (classical)
vacua; for the Sine-Gordon model in the rescaled version
(\ref{eq:genf}) these vacua correspond to constant field
configurations $\phi_0$ being integer multiples of $2\pi$ . In
$\phi^4$-theory there are only two such degenerate vacua at $\phi_0 =
\pm 1$. In any case
these vacua can be classified by integers. A soliton is described by a
field configuration which interpolates between different vacua at
spatial infinity, $i.e$ at $x=-\infty$ and $x=+\infty$.  The basic
soliton is the one interpolating between neighboring vacua. The
associated topological charge, a winding number, is a conserved
quantity. For the basic soliton its value is one. Within the
Sine-Gordon model, according to the bosonization rules
\cite{Col}, this topological charge has to be identified with
fermion number.  Being a conserved quantity, the topological charge
provides a sort of superselection rule with the consequence that the
total Fock space is decomposed into distinct, disconnected Fock spaces
with fixed topological charge. Therefore, ${\cal Z}[j]$ is properly
defined after specifying not only the boundary values of the field in
time but also the boundary values in space, $i.e.$ the topological
charge.

A semiclassical expansion $\phi (x,t) = \phi_0 (x,t) + g\zeta (x,t)$
is based on a classical solution $\phi_0 (x,t) $. In the topologically
trivial sector this classical solution is one of the previously
discussed constants $\phi_0$.  Expanding the Lagrangian in
(\ref{eq:genf}) to second order in the fluctuating field $\zeta (x,t)$
one recovers a Klein-Gordon-type theory with a mass $\mu^2 = V^{(2)}
(\phi_0 )$ given by the second derivative of the potential $V(\phi )$.

In a topologically nontrivial sector the classical solution is no
longer a constant. In the simplest case it is a static solution
$\phi_0 (x)$ obeying the equation
\beq
-\partial_x^2 \phi_0 (x) + V^{(1)} (\phi_0 (x) ) = 0~~.
\label{eq:sol}
\eeq
Here $V^{(1)} (\phi )$ denotes the first derivative of $V(\phi )$ with
respect to $\phi$. The fluctuations $\zeta (x,t)$ on top of such a
soliton can be diagonalized if expanded with respect to a basis of
eigenfunctions $\zeta_n (x)$,
\beq
\zeta (x,t) = a_0 (t) \zeta_0 (x) + \sum_{n=1}^N a_n(t) \zeta_n (x) ~~,
\label{eq:zetaexp}
\eeq
with eigenfunctions $\zeta_n (x)$ and their eigenvalues $\Omega_n$
defined by
\beq
-\partial_x^2 \zeta_n (x) + V^{(2)} (\phi_0 (x) ) \zeta_n (x) =
\Omega_n^2 \zeta_n (x) ~~.
\label{eq:fluc}
\eeq
Usually, there is, in addition to possible bound states, a continuum
of scattering states.  For these states the summation over $n$ has to
be considered a symbolic notation whose meaning is integration.  Due
to the translationally noninvariant soliton solution there is also a
zero mode $\zeta_0 (x) \sim \partial_x
\phi_0 (x)$. It is this mode which restores the spontaneously
broken symmetry.

The physical spectrum of the soliton sector has been discussed in
\cite{GJ} using methods developed for many-body theories \cite{KK}.
The picture emerging from the semiclassical approximation is the
following (seee also \cite{Raj}): the ground state where all
fluctuation modes are carrying out just zero point motion describes a
soliton at rest. Excitation of the zero mode leads to translational
displacement. It is obvious that this mode has to be treated different
simply because its excitation costs no energy and thus a semiclassical
expansion makes little sense. The excitation of a bound state solution
corresponds to an excited soliton whereas the excitation of a
scattering solution corresponds to a soliton-plus-meson state. If the
excited state can decay into soliton-plus-meson states it represents a
resonance.

The usual procedure in connection with (1+1) dimensional soliton
systems introduces a collective coordinate as a time dependent center
of mass coordinate of the soliton field \cite{CL,GS,CC}.  Proper
quantization of the enlarged system requires a constraint in order to
remove the spurious degree of freedom. Such a constraint is nothing
else but a gauge. The physics is completely independent of the choice
of gauge; however, the choice of gauge defines the part of physics
described by the ``collective'' coordinate. If one can solve a theory
exactly the introduction of a collective coordinate would be nothing
else but an academic game. However, the majority of theories requires
an approximate solution. In these cases it can make a big difference
which representation one chooses for the theory. As such a suitable
gauge may give access to completely new approximation schemes.

In connection with soliton models an often used constraint is to
remove the zero mode from the fluctuating field, the so-called
``rigid'' gauge. The disadvantage of this gauge is that it is
formulated in terms of the soliton configuration {\em at rest}. As
such relativistic effects like Lorentz contraction are higher order
effects with the consequence that an expansion in the coupling
constant produces two corrections: Quantum corrections to the soliton
mass and relativistic corrections.  Nevertheless, relativistic
covariance is not lost but just hidden. It requires a summation of
infinitely many Feynman diagrams to show $e.g.$ the relativistic
relation between energy and momentum. On the other hand, for fixed
momentum the soliton is moving with constant velocity; this {\em
classical} part of the collective motion can be accounted for by a
Lorentz-boost to the rest frame of the soliton (see $e.g.$ the first
reference of \cite{GS}). Therefore, as long as {\em fluctuations} of
the collective coordinate are neglected the soliton behaves as a
point-like particle. The question is whether this picture persists if
collective fluctuations are included. As will be shown below this is
not the case.

Another subject discussed in connection with solitonic theories was
the question of meson-soliton vertices. By construction, in a naive
semiclassical expansion there is no linear coupling of the mesonic
fluctuations to the soliton. This was interpreted as the absence of
``Yukawa''-terms, a problem in particular for the Skyrme model
\cite{Sky} in (3+1) dimensions which should describe the nucleon as an
extended particle \cite{Yuk1}. This problem was investigated and
solved using the collective coordinate method \cite{Yuk2}, also in
(1+1) dimensions \cite{Yuk3}. A few years ago, a modification of the
``rigid'' gauge was presented, the so-called ``nonrigid'' gauge
\cite{nrig}, especially designed for the treatment of meson-soliton
scattering and the extraction of meson-soliton vertices.

This nonrigid gauge may serve as an example how a suitable choice of
gauge may simplify calculations. Let me now present a more general
approach making use of the gauge theoretic aspects of collective
coordinate quantization.

\section{Introduction of a collective coordinate}

In the following I shall consider the generating functional
(\ref{eq:genf}) in the soliton sector. Thus,
\beq
{\cal Z} [j] = \langle X_f | {\cal T} \exp
\left(-i\int_{-T'/2}^{T'/2}\! dt\;
\hat{H} \right) | X_i \rangle
\label{eq:defZ}
\eeq
describes the matrix element of the time evolution operator in the
presence of an external source, now between states $\langle X_f|$ and
$|X_i \rangle$. These states specify the ground state in the soliton
sector with the center of mass of the soliton at $X_i$ respectively at
$X_f$. Translational invariance implies that the matrix element
(\ref{eq:defZ}) depends only on the difference $X_f - X_i$. The length
$T'$ of the time interval will finally be taken infinite.

Apart from topological charge there are two other conserved
quantities: total energy and momentum. For the ground state the total
momentum is also the soliton's momentum. For given total momentum the
center-of-mass coordinate of the soliton will therefore classically
describe a trajectory with constant velocity $\beta$. For the matrix
element (\ref{eq:defZ}) this implies that
\beq
\beta = \frac{X_f - X_i}{T} ~~.
\label{eq:beta}
\eeq
Let me account for this situation right from the beginning by choosing
a suitable comoving frame. Here, the external source can serve as a
reference; instead of $j(x,t)$ its arguments are taken to be
\beq
j = j( \bar{x}, \bar{t} ) ~~,
\eeq
with
\beq
\bar{x} = \gamma x +\gamma\beta t +X_0 ~~,~~\bar{t} = \gamma t +
\gamma\beta x  ~~,~~ X_0 = \frac{X_f + X_i}{2}~~.
\label{eq:ref}
\eeq
This implies
\beq
x= \gamma ( \bar{x} - X_0 - \beta \bar{t} )~~, ~~t= \gamma ( \bar{t} -
\beta \bar{x} + \beta X_0 )~~,
\eeq
indicating that the field $\phi (x,t)$ is measured with respect to a
frame moving with velocity $\beta$. As is standard relativistic
notation, $\gamma = 1/\sqrt{1-\beta^2}$.

In order to have a more concrete picture for what ${\cal Z} [j]$
describes recall that in the Sine-Gordon model one can identify the
soliton with a fermion according to the bosonization rules
\cite{Col}; in this case the external source can be viewed as an
electric field.  Even though this is a theory in just one space
dimension this interpretation provides a more familiar physical
picture.

\subsection{Relation to gauge theories}

The introduction of a collective coordinate follows a method described
in \cite{AlDam}. The first step consists in a transformation of the
field $\phi (x,t)$:
\beq
\phi(x,t) = \chi (x-R(t),t)
\label{eq:trans}
\eeq
At this stage, $R(t)$ plays the r\^ole of a parameter, local in time,
describing an arbitrary moving frame.  After a transformation of
coordinates from $(x,t)$ to $(y,t)$ with
\beq
y = x - R(t )~~
\eeq
the generating functional ${\cal Z}[j]$ reads
\bea
{\cal Z} [j] &=& \int\! {\cal D}[\chi ]\; \exp \left(
\frac{i}{g^2}
\int\!  dydt \; {\cal L}'(y,t ) \right) \cr {\cal L}' &=& \frac{1}{2}
\dot{\chi}^2 - \dot{R}\dot{\chi} \chi^{'} -
\frac{1}{2}(1-\dot{R}^2) \chi^{'2}  -  V(\chi ) + j_R \chi
\label{eq:Ztr}
\eea
Here, I introduced the general notation
\beq
\dot{\chi} = \frac{\partial}{\partial t}\chi
(y,t ) \quad , \quad
\chi^{'}  = \frac{\partial}{\partial y} \chi (y,t )
\quad , \quad \dot{R} = \frac{\partial}{\partial t} R(t)
\eeq
$i.e.$ partial derivatives with respect to the first and second
argument. The source which distinguishes between rest frame and moving
frame now depends on $(y,t)$ and $R(t)$ as
\beq
j_R = j( \gamma (y+R) + \gamma\beta t +X_0, \gamma t + \gamma
\beta (y+R))~~.
\label{eq:defjR}
\eeq

In deriving the expression (\ref{eq:Ztr}) I have assumed a
regularization which preserves translational and Lorentz invariance;
otherwise there would be an additional contribution from the Jacobian
of the transformation (\ref{eq:trans}). Another subtlety arises for
$e.g.$ lattice regularization due to different ordering prescriptions;
an example is Weyl ordering in operator formalism corresponding to
mid-point prescription in the discretized functional intergral. In
this case one might get even two-loop corrections from the measure
\cite{2loop}. I will
not go beyond 1-loop, so the latter problem plays no r\^ole here.  A
third point to care about is a possible change of limits of
integration in going from $x$ to $y$ if R(t) reaches infinity
somewhere in the time interval. Later on, $R(t)$ will be quantized
allowing, in principle, highly irregular configurations. At least
within a semiclassical expansion, where a smooth classical path
modified by small fluctuations dominates, these irregular
configurations play no r\^ole. Then, one only has to look at R(t) at
the initial and final times. These are boundary values and will not be
subject to quantum fluctuations. The Lorentz transformation to the
frame moving with velocity $\beta$ is designed to take care of these
boundary values. This means that for an appropriate choice of $\beta $
any $R(t)$ obeys trivial boundary conditions $R(t=-T/2 ) = R(t=+T/2 )
=0$ where $T=T'/\gamma$ is the length of the time interval as measured
in the comoving frame.

The second step towards a quantization of $R(t)$ is to declare it a
dynamical variable, $i.e.$ it becomes a variable of functional
integration. This leads to a trivial local gauge symmetry:
\bea
\chi (y,t) &\to& \chi (y -\alpha (t),t) \cr
R(t) &\to& R(t) - \alpha (t)
\label{eq:gsym}
\eea
The first transformation is equivalent to (\ref{eq:trans}) except that
$R(t)$ is now replaced by $R(t) + \alpha (t)$. Having introduced
functional integration with respect to $R(t)$ the latter can be
shifted by $-\alpha (t)$ precisely compensating the first
transformation in (\ref{eq:gsym}). Obviously, there appear no vector
gauge bosons; only the pure gauge degrees of freedom are involved.

Actually, the generating functional ${\cal Z} [j]$ is completely
independent of $R(t)$ because the latter was introduced by a
transformation of functional integration variables. This independence
allows one to derive Ward identities, in the present case being
related to the conservation of total momentum.  For the integrated
functional $\int\! {\cal D} [R]\; {\cal Z} [j]$ this simply implies an
additional infinite volume factor $\int\! {\cal D}[R]$. This volume
factor has to be cancelled by a suitable constraint, in order to
recover the original generating functional. This constraint can be
regarded as a gauge fixing term. Its choice is almost completely
arbitrary, because it does not influence the physics. It does,
however, influence the amount of physical information carried by
$R(t)$. The simplest case would be the insertion of a
$\delta$-functional requiring $R(t)=0$. This immediately yields the
original representation (\ref{eq:genf}); the variable $R(t)$ would be
completely unphysical, namely zero.

Obviously, one has to find a gauge which allows to identify $R(t)$
with the center-of-mass coordinate of a {\em quantum} soliton. It is
at this point where the equivalence to gauge theories shows its power
for the first time. Recall, that, in principle, any gauge-variant
object can serve as a gauge fixing function.  Therefore, the task is
to find a suitable gauge-variant object. At this point, physical
reasoning gives the answer. Let me consider the conserved quantity
associated with translations, $i.e.$ the total momentum $P(t)$ as
measured in the comoving frame. Due to the presence of $R(t)$ it
becomes modified as
\bea
P (t) &=& -\frac{1}{g^2} \int\! dx\; \phi^{'} (x,t) \dot{\phi} (x,t)
\cr
&=& -\frac{1}{g^2} \int\! dy\; \chi^{'} (y,t) \dot{\chi} (y,t) +
\dot{R} (t) \int\! dy\; \chi^{'2} (y,t)
\label{eq:mom}
\eea
As a physical quantity $P(t)$ is invariant under the gauge
transformation (\ref{eq:gsym}) as can be checked explicitly.  However,
the individual terms in the last expression are gauge-variant. The
first term of the last expression can be interpreted as the momentum
of the field $\chi$ measured with respect to an additional moving
frame now defined by $R(t)$.  Then, the last term is the momentum due
to the motion of this frame itself; it is this term which I want to
identify with the total momentum $P(t)$. The gauge, therefore, has to
be
\beq
G(t) = \frac{1}{g^2} \int\! dy\; \chi^{'} (y,t) \dot{\chi} (y,t) = 0
\label{eq:gauge}
\eeq
which is nothing else but the definition of the center of mass frame.
Note, that this gauge does not refer to any approximate solution of
the underlying field theory.

The gauge fixing procedure is completed once the corresponding
Faddeev-Popov determinant \cite{FP} is found. Here, it is given by
\beq
\det \frac{\delta G(t)}{\delta \alpha (t')} = \det \biggl(
\frac{1}{g^2} \int\! dy\; \chi^{'2} (y,t) \partial_t \delta (t-t')
\biggr)
\label{eq:FPdet}
\eeq
The gauge fixing terms (\ref{eq:gauge}) and (\ref{eq:FPdet}) give
raise to additional terms in the Lagrangian once an auxiliary field
$b(t)$, $i.e.$ a Lagrange-multiplier for the gauge constraint, and
anticommuting ghost fields $c(t),\bar{c} (t)$ for the Faddeev-Popov
determinant are introduced \cite{AlDam}. The generating functional
(\ref{eq:genf}) finally reads
\bea
{\cal Z} [j] &=& \int\! {\cal D}[\chi ]{\cal D}[R]{\cal D}[b] {\cal
D}[c,\bar{c}] \;
\exp\left(\frac{i}{g^2}\int\! dydt\; {\cal L}''(y,t) \right) \cr
{\cal L}'' &=& {\cal L}' + b(t) \dot{\chi} (y,t) \chi^{'} (y,t) +
\bar{c} (t)\chi^{,2} (y,t) \dot{c} (t)
\label{eq:Lgf}
\eea
The gauge symmetry (\ref{eq:gsym}) of the enlarged system is now
broken by the gauge fixing terms. Instead, the functional
(\ref{eq:Lgf}) has a global BRST-symmetry. The corresponding
BRST-transformations are given by
\bea
 \delta_{BRST}\; \chi (y,t) &=& -c(t) \chi^{'} (y,t)\cr
\delta_{BRST}\; R(t) &=& - c(t)\cr \delta_{BRST}\; b(t) &=& 0\cr
\delta_{BRST}\; c(t) &=& 0\cr \delta_{BRST}\; \bar{c} (t) &=& b(t)
\label{eq:BRST}
\eea

This BRST-symmetry is the second advantage of this approach. It can be
used to derive a series of Ward-identities relating certain
n-point-functions of the full theory. This allows to consistently
ensure the constraint in any approximation which is governed by an
expansion parameter. Let me consider the relevance of BRST-invariance
in a more detailed manner: The additional gauge fixing terms in
(\ref{eq:Lgf}) can be written as a BRST-variation
\beq
\int\! dy\; {\cal L}'' = \int\! dy\; {\cal L}' + \delta_{BRST}
(\bar{c} G )
\eeq
Now, let me choose a slightly different gauge $G(t) +\Delta G(t)$.
Gauge invariance of any n-point function to be calculated with the
functional integral (\ref{eq:Lgf}) means
\beq
\Delta \langle {\cal F}_n \rangle = i\langle \int\! dt\;
\delta_{BRST} (\bar{c} \Delta G )\; {\cal F}_n \rangle = i \langle
\int\! dt\; (\bar{c} \Delta G ) \; \delta_{BRST}
({\cal F}_n )\rangle = 0~~.
\label{eq:ginv}
\eeq
Here, $\langle \ldots \rangle$ denotes averaging with respect to the
functional integral (\ref{eq:Lgf}).  In order to fulfill equation
(\ref{eq:ginv}) for arbitrary $\Delta G$ one therefore has the
requirement of BRST-invariance of ${\cal F}_n$.  In the present case
this is trivial if physical quantities are defined in terms of the
original $\phi (x,t)$. This automatically leads to gauge and
BRST-invariant n-point-functions.

However, this is true only for the full theory. What happens if one
approximates the full theory? What about gauge invariance? The answer
can be found repeating the same analysis as in (\ref{eq:ginv}) but now
assuming an approximate Lagrangian ${\cal L}_{app}$. When performing a
``partial integration'' of the BRST-variation as in (\ref{eq:ginv})
one now receives, in general, a contribution from the approximate
Lagrangian. For the full theory this was absent because ${\cal L}''$
was BRST-invariant.  For an approximate Lagrangian to some order $k$
of an expansion one gets instead
\beq
\langle \int\! dt\; (\bar{c} \Delta G) \biggl( \delta_{BRST} {\cal F}_n
+ i \biggl[ \int\! dydt\; \delta_{BRST} {\cal L}_{app}^{(k)}\biggr]
{\cal F}_n \biggr)\rangle = 0 ~~.
\eeq
Gauge invariance to the order $k$ of approximation is now certainly
ensured if
\beq
 \delta_{BRST} {\cal F}_n + i\biggl[ \int\! dydt\; \delta_{BRST} {\cal
L}_{app}^{(k)} \biggr] {\cal F}_n = {\cal O} (k+1)~~.
\eeq
Now, ${\cal F}_n$ can be an arbitrary n-point function and one is
mainly interested in BRST-invariant n-point functions. A general
requirement for a consistent approximation therefore reads
\beq
\delta_{BRST} {\cal L}_{app}^{(k)} = {\cal O} (k+1)~~,
\eeq
$i.e.$ the approximate Lagrangian should be BRST-invariant up to the
order of approximation it governs.

The previous discussion shows that, given a certain approximation
scheme, BRST-invariance guarantees gauge invariant results at any
order. However, this is no statement about the quality of the
approximation. Indeed, in the next subsection I will present a naive
semiclassical expansion also covering the zero-mode. Such a naive
treatment neglects completely the opportunities provided by the
introduction of a collective coordinate. Indeed, one of the great
advantages of an explicit collective coordinate is the access to new
approximation schemes. A first attempt in this direction is presented
in section 4.

Finally a comment on the gauge fixing procedure: the careful reader
may have noticed that the gauge $G(t)$ does not fix all possible
modes; time independent gauge transformations are not gauge fixed and
the Faddeev-Popov determinant would thus be singular due to the
constant mode of the ghost field $c(t)$.  However, a constant $R(t) =
x_0$ is just a translation and thus a symmetry of our theory, suppose
we neglect the external source $j(x,t)$ or treat it as a perturbation.
As a consequence, $R(t)$ is only defined modulo constants. We can
therefore restrict the paths $R(t)$ to nonconstant configurations. The
same applies to the gauge transformations which involve only
nonconstant $\alpha (t)$. Due to BRST variations this means that also
the ghost $c(t)$ is restricted to nonconstant modes. Keeping this in
mind the gauge fixing procedure is complete and no singularities
should arise.

\subsection{A naive semiclassical approximation}

In order to illustrate the use of BRST-symmetry let me now perform a
semiclassical expansion analogous to the one in section 2. The field
$\chi (y,t)$ is divided into a classical and a fluctuating part, the
latter being proportional to $g$ (which is equivalent to
$\sqrt{\hbar}$):
\beq
\chi (y,t) = \chi_0 (y,t) + g\eta (y,t )
\eeq
The fluctuating field obeys trivial boundary conditions, $i.e.$ it
vanishes at spatial infinity as well as at initial and final time.  A
rescaling of the auxiliary fields,
\bea
b(t) &\to& g b(t) \cr
\bar{c} (t) &\to& g \bar{c} (t) \cr
c(t) &\to& g c(t) ~~,
\eea
turns out to be useful. Similarly as for the field $\chi (y,t)$ we
separate $R(t)$ into a classical part and a fluctuating part:
\beq
R(t) = R_0 (t) + g q(t)
\eeq
The BRST transformations of the quantum fields now read:
\bea
 \delta_{BRST}\; \eta (y,t) &=& -c(t) \chi_0^{'} (y,t) - g c(t)
\eta^{'} (y,t) \cr
 \delta_{BRST}\; q(t) &=& - c(t)\cr \delta_{BRST}\; b(t) &=& 0\cr
\delta_{BRST}\; c(t) &=& 0\cr \delta_{BRST}\; \bar{c} (t) &=& b(t)
\label{eq:flBRS}
\eea
Only the variation of $\eta (y,t)$ has a $g$-dependent part.

The semiclassical expansion of the Lagrangian of (\ref{eq:Lgf}) reads
\beq
{\cal L}'' = {\cal L}_0 + g {\cal L}_1 + g^2 {\cal L}_2 + g^3 {\cal
L}_3 + \ldots
\eeq
The classical part is given by ${\cal L}_0$:
\beq
{\cal L}_0 = \frac{1}{2} \dot{\chi}_0^2 - \dot{R}_0 \dot{\chi}_0
\chi_0^{'} -
\frac{1}{2} (1-\dot{R}_0^2) \chi_0^{'2} -  V(\chi_0)
\eeq

The next term, ${\cal L}_1$, is linear in the quantum fluctuations.
Requiring that the linear terms vanish leaves us with
\beq
{\cal L}_1 = j_{R_0} \chi_0
\eeq
and defines the field equation,
\beq
-\ddot{\chi}_0 + \ddot{R}_0 \chi_0^{'} + 2\dot{R}_0
\dot{\chi}_0^{'}+ (1-\dot{R}_0^2) \chi_0^{''} - V^{(1)} (\chi_0 )
= 0~~,
\label{eq:eqm1}
\eeq
as well as the constraint on the classical field $\chi_0 (y,t)$:
\beq
\int\! dy\; \dot{\chi}_0 \chi_0^{'} = 0
\label{eq:eqm2}
\eeq
Note that this set of equations is equivalent to the original field
equation in terms of $\phi (x,t)$; the appearence of an additional
degree of freedom $R_0 (t)$ in (\ref{eq:eqm1}) is compensated for by
the constraint (\ref{eq:eqm2}). From the term linear in $q(t)$ one can
derive another equation describing the conservation of the total
momentum:
\beq
\partial_t \left[ \dot{R}_0 \int\! dy\;
\chi_0^{'2} \right] = 0
\label{eq:eqm3}
\eeq
This equation is, however, not independent of (\ref{eq:eqm1}) and
(\ref{eq:eqm2}).

The last term in the semiclassical expansion I am taking into account
is the one bilinear in the fluctuating fields:
\bea
{\cal L}_2 &=& \frac{1}{2} \dot{\eta}^2 - \dot{R}_0 \dot{\eta}
\eta^{'} -
\frac{1}{2} (1-\dot{R}_0^2 )\eta^{'2} - \frac{1}{2} V^{(2)} (\chi_0)
\eta^2 + j_{R_0} \eta \cr
&& + b\dot{\eta} \chi_0^{'} + b \eta^{'} \dot{\chi}_0 + \bar{c}
\chi_0^{'2} \dot{c} \cr
&& \frac{1}{2} \dot{q}^2 \chi_0^{'2} - \dot{q} ( \dot{\eta}
\chi_0^{'} + \eta^{'} \dot{\chi}_0 - 2\dot{R}_0 \eta^{'}
\chi_0^{'} ) - q j_{R_0}
\chi_0^{'}
\label{eq:fluct}
\eea
Higher order terms in this expansion can be derived in a systematic
manner.

This completes the list of terms which are needed in the following.
The solution of the field equation (\ref{eq:eqm1}) and the constraint
(\ref{eq:eqm2}) as well as the diagonalization of ${\cal L}_2$
yielding the fluctuation normal modes can now be performed
analytically. It remains to be shown that the approximation by ${\cal
L}_2$ is BRST-invariant up to this order.  This is easily done. Taking
into account the field equation (\ref{eq:eqm1}) and the constraint
(\ref{eq:eqm2}) the BRST-variation of ${\cal L}_2$ is of order ${\cal
O} (g)$ and thus negligible in this approximation. This guarantees
gauge invariance up to ${\cal O} (g)$.

The leading order gauge constraint (\ref{eq:eqm2}) is easily fulfilled
by a static configuration
\beq
\chi_0 (y,t) = \chi_0 (y)~~.
\eeq
This ansatz and equation (\ref{eq:eqm3}) imply a constant velocity
$\dot{R}_0$. Keeping in mind that the moving frame was chosen such
that $R (t)$, and therefore also $R_0 (t)$, satisfies trivial boundary
conditions one can take $\dot{R}_0 =0$ and, due to translational
invariance, also $R_0 =0$. The equation for $\chi_0 (y)$ now reads
\beq
\chi_0^{''} (y) - V^{(1)} (\chi_0 (y) ) = 0
\eeq
This equation is just the classical field equation for a soliton.  Its
solution therefore is
\beq
\chi_0 (y) = \phi_0 (y) ~~.
\eeq
In the reference frame, $\chi_0 (y) = \phi_0 (\gamma (\bar{x} - X_0 -
\beta \bar{t})$ appears as a  Lorentz-boosted classical soliton. Its
rest mass is given by
\beq
M_0 = \int\! dy\; \chi_0^{'2} (y) ~~.
\label{eq:defM0}
\eeq

Let me now turn to the fluctuation modes. Their Lagrangian ${\cal
L}_2$, taking into account the properties of the classical solutions
$R_0 (t)$ and $\chi_0 (y)$, reads:
\bea
{\cal L}_2 &=& \frac{1}{2} \dot{\eta}^2 -
\frac{1}{2} \eta^{'2} - \frac{1}{2} V^{(2)} (\phi_0) \eta^2 + j\eta \cr
&& + b \dot{\eta} \phi_0^{'} + \bar{c} \phi_0^{'2} \dot{c}\cr &&
+\frac{1}{2} \dot{q}^2 \phi_0^{'2} - \dot{q} \dot{\eta} \phi_0^{'} -
qj\phi_0^{'}
\label{eq:L2}
\eea
As in (\ref{eq:zetaexp}) one expands $\eta (y ,t )$ in a basis of
eigenfunctions $\zeta_n (y )$ of the hermitian operator
$-\partial_{y}^2 + V^{(2)} (\phi_0 (y ))$, with coefficients $a_n (t
)$. The functions $\zeta_n (y )$ are precisely the eigenfunctions of
(\ref{eq:fluc}), now in the {\it rest frame of the moving soliton}
$\phi_0 (y )$. The zero mode wave function $\zeta_0 (y )$ is related
to the classical soliton solution $\phi_0 (y )$ through
\beq
\zeta_0 (y ) = \frac{\phi_0^{'} (y )}{\sqrt{M_0}}  ~~.
\eeq

The Lagrange function $L_2 = \int\! dy\; {\cal L}_2$ describes a set
of oscillators with coordinates $a_n (t ), n\geq 1$, the collective
fluctuation mode $q(t)$ and a set of spurious zero energy modes:
\bea
L_2 &=& \frac{1}{2} \sum_{n\neq 0} \biggl( \dot{a}_n^2 -
\Omega_n^2 a_n^2 \biggr) + \sum_{n\neq 0} a_n \int\! dy \; j \zeta_n
\cr
&& + \frac{M_0}{2} \dot{q}^2 - \sqrt{M_0} \dot{q} \dot{a}_0 - q
\sqrt{M_0}
\int\! dy\; j\zeta_0 \cr
&& + \frac{1}{2} \dot{a}_0^2 + \sqrt{M_0} b \dot{a}_0 + M_0
\bar{c}
\dot{c}
\label{eq:diagL2}
\eea
The coupling to $b$ essentially sets $\dot{a}_0 = 0$ in any n-point
function which does not depend on $b$. Constant $a_0$ should be
excluded as a quantum degree of freedom because it is related to
translational symmetry.  We can therefore neglect $a_0$ and find for
the physically relevant part the usual fluctuation modes except that
now $a_0$ is replaced by $\sqrt{M_0} q$.

It is not surprising that up to this order the Lagrange function
(\ref{eq:diagL2}) coincides with expressions using different gauges,
as for example the rigid gauge \cite{Sol}. Using the same
semiclassical expansion and requiring gauge invariance should lead to
the same results. Indeed, the effective collective Lagrange function
in the rest frame of the soliton reads
\beq
L_{eff} = - \frac{1}{g^2} M_0 - \sum_{n\neq 0}
\frac{\Omega_n}{2} + \frac{M_0}{2} \dot{q}^2 + {\cal O} (g^2)~~.
\eeq
This is the usual expansion to be found in $e.g.$ \cite{Sol,Raj}.  It
is clear that counting $q(t)$ as of ${\cal O}(g)$ requires the
inclusion of terms of order ${\cal O} (g^2)$ in order to recover the
equivalence of rest mass and kinetic mass up to ${\cal O} (1)$. This
is another hint that the collective coordinate and therefore also the
zero mode has to be treated different.

\section{A modified semiclassical expansion}

The semiclassical approximation of the previous subsection has to be
modified such that $q(t)$ no longer counts as ${\cal O}(g)$. A glance
at the BRST-transformations (\ref{eq:flBRS}) tells us that also $a_0
(t)$ has to be treated different. Actually, one has to solve the gauge
constraint for $a_0 (t)$. For that purpose let me write down the gauge
$G(t)$ in terms of the $a_n (t)$ separating the terms containing $a_0
(t)$:
\bea
G(t) &=& \dot{a}_0 \sqrt{M_0} + g \dot{a}_0 \sum_{n\neq 0} a_n
\biggl\{ \zeta_0 \zeta_n^{'}\biggr\}  - g a_0 \sum_{n\neq 0} \dot{a}_n
\biggl\{ \zeta_0 \zeta_n^{'}\biggr\} \cr
&& + g \sum_{m,n\neq 0} \dot{a}_m a_n \biggl\{ \zeta_m \zeta_n^{'}
\biggr\}\cr
&=& \Bigl( \sqrt{M_0} \partial_t + {\cal O}(g) \Bigr) a_0 + {\cal
O}(g)
\label{eq:a0}
\eea
Here, I have introduced the notation
\beq
\biggl\{ abc \biggr\} = \int\! dy \; a(y) b(y) c(y)~~.
\eeq

{}From the last line of (\ref{eq:a0}) written in compact notation it
is obvious that one can solve the gauge constraint for $a_0$, at least
formally. First of all, it turns out that $a_0 = {\cal O}(g)$.
Secondly, in order to avoid additional factors in the measure when
integrating out $b$ and $a_0$ one has to transform
\beq
\bar{c} \to \bar{c} \Bigl( \sqrt{M_0} \partial_t + {\cal O}(g)
\Bigr)~~.
\eeq
Performing the integration over $b$ and $a_0$ then leads to the
following Lagrange function:
\bea
L' &=& -\frac{M_0}{g^2} + \frac{1}{2} \sum_{n\neq 0} \biggl(
\dot{a}_n^2  - \Omega_n^2 a_n^2 \biggr) + \sum_{n\neq 0} a_n \biggl\{
j \zeta_n \biggr\} +\sqrt{M_0} \bar{c}c \cr && + \frac{1}{2g^2}
\dot{q}^2 \Bigl( M_0 + 2g\sqrt{M_0} \sum_{n\neq 0} a_n \biggl\{
\zeta_n^{'} \zeta_0 \biggr\} + {\cal O}(g^2) \Bigr) +
\frac{1}{g} \biggl\{ j_q \phi_0 \biggr\}\cr
&& + {\cal O}(g)
\label{eq:L2'}
\eea
Here, the factor $1/g^2$ in front of the action in (\ref{eq:Lgf}) is
included in (\ref{eq:L2'}). I have written down all terms relevant for
the construction of an effective collective Lagrange function at
leading order tree level. Here, tree level is meant only with respect
to the $a_n$-modes with $n\neq 0$, $i.e.$ I consider the interaction
of $q(t)$ with the $a_n (t)$ excluding loops. Only the leading order
terms in $g$ are considered.

Let me proceed with such a construction. The only term which has to be
accounted for is the coupling of $\dot{q}^2$ to the $a_n$.  Including
this term one gets
\bea
L_{eff} &=& - \frac{M_0}{g^2} + \frac{M_0}{2g^2} \dot{q}^2 (t)\cr && -
\frac{M_0}{2g^2} \dot{q}^2 (t) \int\! dt'\; \dot{q}^2 (t')
\sum_{n\neq 0} \int\!\frac{d\omega}{2\pi} \; \frac{e^{-i\omega
(t-t')}}{\omega^2 - \Omega_n^2 +i\epsilon} \biggl\{ \zeta_0
\zeta_n^{'} \biggr\}^2\cr
&& + \frac{1}{g} \biggl\{ j_q \phi_0 \biggr\} + {\cal O} (\dot{q}^6
,g^0)~~.
\eea
This is a nonlocal expression. However, recall that $q(t)$ describes
the fluctuations around the classical trajectory.  There are certainly
cases where these fluctuations can be regarded small. This means that
an expansion in time derivatives makes sense.  Therefore, expanding in
$\omega^2/\Omega^2_n$ and using the relation (see second reference of
\cite{GS})
\beq
\biggl\{ \zeta_0 \zeta_n^{'} \biggr\} = \frac{\Omega_n^2}{2} \biggl\{
\zeta_0 y \zeta_n \biggr\}
\eeq
one arrives at
\beq
L_{eff} = - \frac{M_0}{g^2} + \frac{M_0}{2g^2} \dot{q}^2 (t) +
\frac{M_0}{8g^2} \dot{q}^4 (t) + \frac{M_0}{2g^2} r_{ms}^2 \dot{q}^2
(t) \ddot{q}^2 (t) + \ldots + \frac{1}{g} \biggl\{ j_q \phi_0 \biggr\}
\label{eq:Leff}
\eeq
Here, the dots denote terms of ${\cal O} (g^0)$ or of higher time
derivatives. The expression $r_{ms}^2$ denotes the mean square radius
of the mass distribution of the soliton, $i.e.$
\beq
r_{ms}^2 = \int\! dy \; y^2 \zeta_0^2 (y)~~.
\eeq

The first three terms in (\ref{eq:Leff}) can be interpreted as
resulting from a Taylor expansion of $\sqrt{1-\dot{q}^2}$. Indeed, if
we consider the action of a relativistic point-like particle with mass
$M$,
\beq
{\cal S} = - \int\! ds\; M \sqrt{\dot{x}_\mu (s) \dot{x}^\mu (s)} ~~,
\label{eq:Spoint}
\eeq
and introduce the fluctuation $q(t)$ {\em in the moving frame} as
\beq
x^0 = \gamma t + \gamma \beta q(t) \quad ,\quad x^1 = \gamma \beta t +
\gamma q (t)
\eeq
we find for the first three terms
\beq
{\cal S} \simeq \int\! dt \; \biggl( - M + \frac{M}{2} \dot{q}^2 (t)+
\frac{M}{8} \dot{q}^4 (t) + {\cal O} (\dot{q}^6 ) \biggr)~~.
\label{eq:Sfluct}
\eeq
This expansion looks like a nonrelativistic expansion. It should be
noted, however, that $t$ and $q(t)$ denote the time respectively the
fluctuation around the classical straight line path as measured in its
rest frame.

It is obvious that the action of a point-like particle depends only on
the velocity of the fluctuations. The fourth term in (\ref{eq:Leff})
uncovers a substantial difference in the quantum mechanics of a
soliton as compared to the one of a point-like particle. This is
almost evident due the appearence of the mean-square radius
$r_{ms}^2$.  The origin of this term lies in the interaction of the
collective fluctuation $q(t)$ with the intrinsic field $\chi (y,t)$.
Indeed, treating the {\em acceleration} $\ddot{q}(t)$ as
perturbatively small, the classical field equation now including the
$q(t)$-terms has the soliton solution
\beq
\chi_0 (y,t) = \phi_0 (\gamma (t) y)
\label{eq:adiab}
\eeq
with
\beq
\gamma (t) = \frac{1}{\sqrt{1-\dot{q}^2 (t)}}~~.
\eeq
Note that this solution obeys the constraint,
\beq
\int\! dy \; \dot{\chi}_0 (y,t) \chi_0^{'} (y,t) = \gamma^2 (t)
\dot{q} (t) \ddot{q} (t) \int \! dz \; z \phi^{'2} (z)~~,
\eeq
suppose $\phi_0^{'2} (z)$ is an even function of $z$. This is the case
for a soliton solution centered at $z=0$.

With the field (\ref{eq:adiab}) the classical action reads
\bea
S_{cl} &=& \frac{1}{g^2} \int \! dtdy\; \left( \frac{1}{2}
\dot{\chi}_0^2 - \frac{1}{2} (1-\dot{q}^2) \chi_0^{'2} - V(\chi_0
)\right) \cr &=& \int\! dt\; \left( \frac{M_0}{2g^2} r_{ms}^2
\gamma^3 (t) \dot{q}^2 (t) \ddot{q}^2 (t) - \sqrt{1- \dot{q}^2
(t)}
\frac{M_0}{g^2} \right)~~.
\label{eq:Scl}
\eea
This derivation demonstrates that in an adiabatic limit treating the
velocity $\dot{q} (t)$ as approximately constant the correct Lorentz
covariant expression for the action of a point-like particle arises.
The corrections to this picture are due to nonadiabatic terms. In
particular, the kinetic term $\dot{\chi}_0^2$ is the origin of the
additional term depending on $\ddot{q} (t)$. As such this term can be
understood as a feedback of the intrinsic field $\chi_0 (y,t)$ due to
the accelerated collective motion along the path $q(t)$. In the
adiabatic approximation this feedback is a time dependent
Lorentz-contraction.

What is the consequence of this feedback? Consider the situation where
a weak external force induces periodic motion $q(t)$ with a frequency
$\omega$. The response $q(t)$ to such a force is given by the
propagator or Green's function of $q(t)$. Already from the classical
action (\ref{eq:Scl}) it is clear that accelerated motion gets an
additional suppression from the $\ddot{q}$-term as compared to a
point-like particle. Indeed, the feedback of the intrinsic field
produces a sort of frequency cutoff; for periodic acceleration with a
frequency above this cutoff the soliton may eventually respond weaker
than a point-like particle. Of course, there may also be a resonance
at higher frequencies. Let me exclude this possibility for the moment.

I now present a rough estimate of such a cutoff: For that purpose I
keep only the term quadratic in $\dot{q} (t)$ and the term containing
$\ddot{q}$; the latter being fourth order in $q(t)$ is treated in a
Hartree-type approximation. The Green's function for $q(t)$ now reads
\beq
\langle q(t) q(t') \rangle = i\int\! \frac{d\omega}{2\pi}\;
\frac{e^{-i\omega (t-t')}}{\omega^2 + 2r_{ms}^2 \langle \dot{q}^2
\rangle \omega^4}~~.
\eeq
Instead of performing a selfconsistent approximation let me assume for
convenience that $\langle \dot{q}^2 \rangle = 1/2$; this accounts for
the fact that we are dealing with a relativistic theory and therefore
$\langle \dot{q}^2 \rangle $ should be less than one. In contrast to
the Green's function of a point-like particle\footnote{In the present
approximation a nonrelativistic particle.} an additional factor
$(1+r_{ms}^2\omega^2)$ appears in the denominator leading to a
$1/\omega^4$ behaviour at high frequencies. The transition from the
$1/\omega^2$- to the $1/\omega^4$-behaviour takes place at
\beq
\omega^2 r_{ms}^2 \sim 1~~.
\eeq
The cutoff is thus given roughly by the condition that the
corresponding time period of the external perturbation becomes smaller
than the time needed to transfer information across the soliton. As
such this additional suppression of the response at high frequencies
is a typical retardation effect and probably related to form factors
for elastic scattering.

This derivation of an estimate of the intrinsic cutoff should not be
taken too serious because higher order effects have been neglected.
Nevertheless, the fact that the internal structure of the soliton has
an effect on its {\em quantized} motion is beyond any doubt. Also the
time scale of these effects must be related to the extension of the
soliton, simply due to the familiar uncertainty relation.

\section{Soliton-meson vertices and the coupling to an external source}

As a last topic I consider the question of whether there is $e.g.$ a
sort of Yukawa coupling, $i.e.$ a soliton-meson-soliton vertex.  This
question originates from skyrmion physics \cite{Yuk1}; the classical
skyrmion, a soliton in (3+1) dimensions, has, by definition, no linear
coupling to the fluctuation modes, similar as in the present
(1+1)-dimensional case.  This was interpreted as the absence of a
Yukawa coupling. The resolution of this problem is intimately
connected to the treatment of collective coordinates
\cite{Yuk2,Yuk3}.  With the previous results I will now present
another explanation of this problem.

Recall that the spectrum in the one-soliton sector was discovered
considering the {\em ground state} expectation value of the time
evolution operator in this sector. This ground state consists of a
soliton moving with constant velocity with no mesonic fluctuation
modes excited. A Yukawa coupling cannot show up in first order
perturbation theory for the ground state expectation value, simply
because it changes meson number. It will, however, show up in first
order perturbation theory for an off-diagonal matrix element where
either the initial or the final state contains one meson.  The
corresponding form factors have been discussed by Goldstone and Jackiw
\cite{GJ} and were found to be given by the normal mode wave functions
$\zeta_n (x)$.

How could one extract such a coupling in the present approach? It is
clear that one has to prepare an initial or final state containing one
meson. This meson will propagate forward respectively backward in time
until it will be absorbed due to the meson field operator in the
Yukawa coupling. In the present case, the meson field reads
\bea
\phi (\bar{x} ,\bar{t} ) &=& \phi_0 (\gamma \bar{x} - \gamma X_0 -
\gamma \beta \bar{t} )\cr
&& + g \sum_{n\neq 0} a_n (\gamma \bar{t} - \gamma
\beta \bar{x} + \gamma \beta X_0 ) \zeta_n ( \gamma \bar{x} - \gamma
X_0 - \gamma \beta \bar{t} ) \cr &&+ {\cal O} (g^2)~~,
\label{eq:phi}
\eea
expressed in terms of the laboratory frame coordinates (\ref{eq:ref})
and using the solution of the field equation. Note that the constraint
allows to replace the zero mode $a_0 (t)$ by terms of order ${\cal O}
(g)$. A meson can be absorbed or emitted by this field with an
amplitude given by $\zeta_n (\gamma \bar{x} - \gamma X_0 - \gamma
\beta \bar{t} ) $. The corresponding form factor in the Breit frame is
then given by
\beq
G_n (p) = \frac{1}{\sqrt{2\Omega_n}} \int\! dX_0 e^{-ipX_0}
\zeta_n (-X_0)~~.
\label{eq:ff}
\eeq
Here, the Breit frame is essentially the rest frame of the soliton.
The factor $1/\sqrt{2\Omega_n}$ is just the coefficient which appears
when the ``coordinate'' $a_n (t)$ is expressed in terms of cration and
annihilation operators, respectively their corresponding c-number
analogues in the functional integral.  Result (\ref{eq:ff})
corresponds to the expression derived in
\cite{GJ}.

Now it is easy to analyse the effect of an external source coupling to
$\phi (\bar{x},\bar{t})$. According to (\ref{eq:phi}) such a coupling
creates and absorbs mesons leading to resonance or scattering states.
In addition, there is a coupling of $j(\bar{x},\bar{t})$ to the
classical soliton solution. It measures, the additional interaction
energy. But there is still another effect. In the presence of an
external source the soliton will be accelerated. This can be seen from
(\ref{eq:eqm3}) now also including the external source. The modified
equation for $R_0 (t)$ is
\beq
-M_0 \ddot{R}_0 (t) = \sqrt{M_0} \int\! dy\; j_{R_0} \zeta_0 (y)
\eeq
where the source is treated as a small perturbation and changes in the
mass of the soliton due to excitations have been neglected.  The
zero-mode wave function projects on that part of the external source
which is responsable for an acceleration of the soliton, at least in a
perturbative sense. A nonperturbative treatment requires a presumably
numerical solution of the field euqation and the constraint in the
presence of an external source.

\section{Summary, discussion and outlook}

One aim of this paper has been to present another variation of
introducing collective coordinates with special emphasis on gauge and
BRST-symmetry aspects. I have presented a gauge based on a physical
picture and without any reference to a classical soliton solution. The
result has been a coupled set of equations for the classical
collective coordinate and the soliton field. Lorentz covariance has
been built in right from the beginning introducing the collective
coordinate in a convenient moving frame. This procedure has been
justified by the classical solution $\ddot{R}_0 (t) =0$ together with
the boundary values implying $R_0 (t) =0$.  The second aim was to
emphazise the special r\^ole of the collective coordinate and to work
out the relation to the quantum mechanics of a relativistic point-like
particle.  Deviations from this picture manifest themselves as
additional terms with higher time derivatives in the effective theory
of collective fluctuations. They represent retardation effects and
imply an intrinsic cutoff in the collective response related to the
dimension of the soliton, eventually also a resonant behaviour.
Soliton-meson vertices have been discussed and corresponding form
factors have been derived in accordance with ref.\cite{GJ}.  Finally,
the effect of external sources has been considered. For a weak
external source its effect to leading order has been shown to be
twofold: acceleration of the whole system, $i.e.$ collective
excitation, and internal excitation.

The fact that one can extract collective coordinates analytically and
thereby eventually remove infrared problems from the ``intrinsic''
system (which, of course, depends on the gauge) allows a combination
of numerical and analytical methods, at least in principle. In cases,
where a semiclassical expansion or any other analytical approach for
the intrinsic system fails, one has to use numerical schemes like
Monte-Carlo simulations. It may turn out that one has to choose
appropriate gauges different from the present one; I have already
mentioned that there might be differences in connection with certain
regularization schemes.

I have choosen a soliton theory in (1+1) dimensions for the discussion
of collective coordinate quantization mainly because of its relative
simplicity.  This does not mean that the physical situation is simple.
Indeed, it is precisely the interpretation of solitons as extended
particles with internal structure which makes them so attractive and
also nontrivial.  An application of the present approach to any field
theory which admits symmetry-breaking mean-field solutions\footnote{A
standard situation in e.g. nuclear physics.} seems to be
straightforward. I am not claiming that I have found the ultimate
gauge in connection with collective coordinate quantization even if
this gauge may appear very appealing to some readers. Depending on the
aims different choices may be more advantageous. After all, any gauge
is as good as the other as concerns the physics. The difference is
measured by the calculational effort.

It is worthwile to discuss possible consequences of the results
presented above. In particular, the retardation effects modifying the
behaviour of the collective fluctuations as compared to a point-like
particle may also play a r\^ole when calculating quantum corrections
to the soliton energy. The collective zero-point motion has to be
taken into account and may very well modify existing calculations.

The present approach also allows to calculate the effects of strong
external fields, at least numerically, and still distinguish between
collective and intrinsic phenomena. This is achieved already at the
classical level due to a coupled set of equations for $R_0 (t)$ and
$\chi_0 (y,t)$.

Let me finally consider implications of the previous analysis for
soliton models in (3+1) dimensions. As long as translational motion is
concerned one can expect a similar behaviour. This means that the
quantization of translational motion for a moving soliton can be
carried out in its rest frame. The results are then transformed to the
reference frame by a Lorentz-boost.

The fluctuations of the collective motion now also involve rotations.
Let us restrict considerations to the case where $\vec{R} (t)$ always
points in the same direction. Then, possible differences may still
occur due to the presence of higher derivative terms as in $e.g.$ the
Skyrme-Lagrangian\footnote{This question was brought to my attention
by R. Alkofer}. As long as these higher derivative terms lead to just
higher powers of $\dot{\vec{R}}$ the situation is similar to the
present case, at least at tree level: Recall that due to
Lorentz-invariance of the action all four-derivatives appear
contracted and therefore, after introducing $\vec{R}(t)$, spatial
derivatives appear always in the combination
\beq
\vec{\nabla} F \vec{\nabla} G - (\dot{\vec{R}} \vec{\nabla}) F
(\dot{\vec{R}} \vec{\nabla}) G~~,
\eeq
with $F,G$ field dependent quantities. Therefore, one can repeat the
same arguments leading to the effective action for the collective
fluctuation $\vec{q}(t)$, in particular in the adiabatic limit.

One should also expect terms depending on $\ddot{\vec{q}} (t)$.  Such
terms can arise already before solving the field equations due to
terms in the Lagrangian where the pion field appears with second time
derivatives or even higher. The lesson of the present paper is that
such terms indicate some intrinsic structure. The origin of these
terms in the Lagrangian can be understood if one recalls that these
Lagrangians are usually itself {\em effective} Lagrangians; this means
that they are in some way derived from an underlying theory by
integrating out short distance degrees of freedom, in an analogous way
as for the derivation of the effective Lagrange function
(\ref{eq:Leff}). In a gradient expansion the, in principle, nonlocal
structure of the effective Lagrangian is traded with terms containing
second and higher derivatives acting on the fields. There is already
some intrinsic structure due to the short distance degrees of freedom
which have been integrated out. The relevant length scale has to be
smaller than the extension of the soliton; otherwise, the effective
Lagrangian wouldn't make sense.

The discussion of soliton-meson vertices seems to be applicable in
(3+1) dimensions as well. Of course, the number of vertices will be
much larger due to the richer set of internal quantum numbers like
spin, isospin etc..

I would like to acknowledge discussions with N.N. Scoccola at the
Niels-Bohr-Institute in Copenhagen initiating this paper. Part of this
work has been carried out at Technical University Munich.  Finishing
this work would have been impossible without the support of the theory
group at GSI.

\bibliographystyle{unsrt}

\end{document}